# RADIATION MEASUREMENTS AT THE CAMPUS OF FUKUSHIMA MEDICAL UNIVERSITY THROUGH THE 2011 OFF THE PACIFIC COAST OF TOHOKU EARTHQUAKE AND SUBSEQUENT NUCLEAR POWER PLANT CRISIS

TSUNEO KOBAYASHI [1]

[1] Department of Natural Sciences (Physics), School of Medicine, Fukushima Medical University, Fukushima, Japan

**Abstract:** An earthquake, Tohoku region Pacific Coast earthquake, occurred on the 11th of March, 2011, and subsequent Fukushima nuclear power plant accidents have been stirring natural radiation around the author's office in Fukushima Medical University (FMU). FMU is located in Fukushima city, and is 57 km (35 miles) away from northwest of the Fukushima Daiichi nuclear power plant. This paper presents three types of radiation survey undertaken through the unprecedented accidents at the campus and the hospital of FMU. First, a group of interested people immediately began radiation surveillance; the group members were assembled from the faculty members of "Life Sciences and Social Medicine" and "Human and Natural Sciences." Second, the present author, regardless of the earthquake, had serially observed natural radiations such as gamma radiation in air with NaI scintillation counter, atmospheric radon with Lucas cell, and second cosmic rays with NaI scintillation. Gamma radiation indicated most drastic change, i.e., peak value (9.3 times usual level) appeared on March 16, and decreased to 1.7 times usual level after two months. A nonlinear least squares regression to this decreasing data gave short half-life of 3.6 days and long half-life of 181 days. These two apparent half-lives are attributed to two groups of radioisotopes, i.e., short half-life one of I-131 and long half-life ones of Cs-134, Cs-137 and Sr-90. Also, atmospheric radon concentration became high since a stop of ventilation, while second cosmic rays did not show any response. Third, late April, 2011, a team of radiation dosimetry under the direct control of Dean, School of Medicine, was established for the continuation of radiation survey in the campus and the hospital of Fukushima Medical University.

**Running Head:**   RADIATION MEASUREMENTS AT THE CAMPUS OF FMU

**INTRODUCTION**

Fukushima Medical University (later often referred to as FMU) is located in the northeastern region of Japan (37°45'N, 140°28'E, 67.4 m above sea level), and is 57 km (35 miles) away from northwest of the Fukushima Daiichi nuclear power plant. The Tohoku region Pacific Coast earthquake occurred on 11 March, 2011, and Fukushima nuclear power plant accidents broke out subsequently.

The present paper reports three groups of radiation measurement performed through the unexampled accidents. First, immediately after the earthquake, in order to investigate the levels of radiations outside and inside the campus, Associate Dean, School of Medicine, Professor Hiroyuki Yaginuma assembled a group of interested people from the faculty members of "Life Sciences and Social Medicine (basic medical sciences)" and "Human and Natural Sciences (liberal arts course)." Their hard-working efforts gave valuable information about radiation safety for many staff and patients in the hospital of FMU and the staff and students of FMU campus.

Second, the present author had been measuring several natural radiations[1] around his office from September in 2010. The natural radiations under the serial measurement were: gamma radiation in air with NaI scintillation counter, atmospheric radon with Lucas cell, and second cosmic rays with another NaI scintillation counter. Amongst the results of serial observation, gamma radiation showed the most drastic change, i.e., peak value of 9.3 times as usual level occurred on March 16, and exponentially decreased to 1.5 times of usual level after five months. A nonlinear least squares regression to these data indicated short half-life of 3.6 days and long half-life of 181 days. The first apparently short half-life (later referred as HL) is attributed to the existence of I-131 (HL: 8 days), while the second long HL may be contributions from nuclides of Cs-134 (HL: 2 years), Cs-137 (HL: 30 years) and Sr-90 (HL: 28.1 years). Also, the atmospheric radon concentration at the other place became high because of a stop of ventilation. Atmospheric radon at other places and second cosmic rays did not show any distinct response.

Third, several weeks after the accidents, a team of radiation dosimetry under the direct control of Dean, School of Medicine, Prof. Hitoshi Ohto, was established for the continuation of radiation survey in FMU.

**MATERIAL AND METHODS**

*1. Radiation surveillance right after the magnitude 9.0 earthquake*

Associate Dean, School of Medicine, Professor H. Yaginuma supervised radiation surveillance groups of interested people mentioned above. An NaI(Tl) scintillation counter surveyed radiation in several places in Fukushima Medical University hospital, i.e, ICU, NICU and pediatric ward. In the early stage, Japanese Self-Defense Force officials conducted patients screening with Geiger counters at the entrance of hospital with the help of this surveillance team.

*2. Serial natural-radiation measurements from September 2010*

NaI scintillation counter

A 3"x3" NaI(Tl) scintillation detector (Teledyne S-1212-T, 7% resolution for Cs-137) was observing gamma radiation in air at the author's office since October 2010. Every four hour counting data was stored in a personal computer. The present report discusses only gross dose rate expressed as a unit of cps. The office room was on the fourth floor of the five storied concrete building built in 1988.

Radon detector

Passive type detectors (Pylon, AB-5) had been measuring atmospheric radon concentration at three places, i.e., the author's office room, students' lab and the Radioisotope Center). Another active type radon detector (Durridge, RAD7) had been detecting atmospheric radon in the author's room. Both types of detector were acquiring every one-hour data and stored in memories and/or printed out to papers. All these detectors safely continued measurements in spite of the magnitude 9.0 earthquake.

Second cosmic rays

For the observation of cosmic rays, a 1"x1" NaI(Tl) scintillation detector (Harshaw 905-3, 7% resolution for Cs-137) had been counting radiation whose energy is over 3 MeV. These data were also stored in a personal computer.

*3. The team of radiation dosimetry under the direct control of Dean, School of Medicine*

The team of radiation measurement under the direct control of Dean, School of Medicine, used mainly NaI(Tl) scintillation counter for hospital and campus surveillance, and again Geiger counters checked mats in entrance hall and shoes-soles of students after exercise of sports in the ground of the campus.

**RESULTS**

*1. Radiation surveillance right after the magnitude 9.0 earthquake*

From the beginning stage of the power plant crisis, Associate Dean, School of Medicine, Professor H. Yaginuma emitted the information of radiation surveillance, acquired from his teams, to FMU staff. In particular, outdoor gamma radiation results were reported on a bulletin board up to the present. Maximum of the observed value was 11.9 µSv/h at 11:30 on the 16th of March, 2011. The outdoor gamma values are around 0.4 µSv/h nowadays.

Ward surveys were not announced officially, however ward members were able to have no worries about indoor radiation.

Patients screening services at the entrance of the hospital continued until the 25th of March, 2011. People whose cps exceeded 10,000 cpm were required decontamination at the separate 'decontamination tent.' The maximum value found was 100,000 cpm. All members of FMU appreciate the aid of the staff of Japanese Self-Defense Force officials.

*2. Serial natural-radiation measurements from September 2010*

Gamma radiation in air

Figure 1 shows gamma radiations with the NaI(Tl) scintillation detector. Just when the magnitude 9.0 earthquake happened, there was no change for the counts. As indicated in the figure, radiation dose reading began to elevate about 18:00 on the 15th of April and reached the maximum value of 9.3 times as usual values. This sudden increase was attributed to the hydrogen explosion at the nuclear power plant. Thereafter, the counts seemed to attenuate exponentially. This early exponential attenuation was expressed roughly 3 days of HL that was attributed to the existence of I-131 (HL: 8.06 days). This early environmental HL of 3 days was reasonably shorter than the physical HL of I-131.

After one month has passed, semi-log plot of the cps versus time did not fit to a single line and the apparent HL became longer and longer, indicating the appearance of the second long HL. Thus, the author tried a model equation of CPS= a $2^{-t/Ta}$ + b $2^{-t/Tb}$, where Ta is the short HL and Tb is the longer HL. Nonlinear least squares regression, using a command, nls, of S-PLUS[2] or R[3], obtained the value of Ta = 3.63 ± 0.02 days and Tb = 181 ± 5 days. This second longer HL increased longer and longer afterwards and reached 181 days after about five months (144 days after the peak value was observed). This long HL might be contributions of newly supplied radioisotopes from nuclear reactor or environmentally accumulated long HL isotopes such as Cs-134, Cs-137 and Sr-90. The value of b/(a + b) was 0.13 that means long HL radioisotopes in FMU were 13% at the first attack of the explosion of the reactor to Fukushima.

Atmospheric radon

Figure 2 shows the change of atmospheric radon in one room of the Radioisotope Center. This Center is normally ventilated extremely because of the prevention of non-sealed radioisotopes. This extreme ventilation stopped when the earthquake broke out and RI users were immediately prohibited to enter. The stop of ventilation naturally raised atmospheric radon concentration as indicated in the Figure 2.

Usual radon level was quite low and less than the lower detection limit of the device (5 Bq/m$^3$), while after the ventilation stopped, it raised as high as 250 Bq/ m$^3$) that exceeds intervention level of the U.S.A.(150 Bq/m$^3$) or Europe (200 Bq/m$^3$). The origin of the elevated radon might be the thick concrete wall of the room that was devised to handle sealed radiation sources. Although the level became high, there was no problem of radiation protection because people were inhibited to enter during the accident.

Radon level at other places also became slightly high after the earthquake, and again the ventilation stopped for about 10 days. However, the level did not exceed usually observed maximum level since these rooms have not so thick concrete as RI center's room.

Second cosmic rays

Second cosmic rays showed a little decrease and growth through the earthquake. However, these changes were explained with the contrary change of the atmospheric pressure; second cosmic rays decrease when the atmospheric pressure increases (thicker air disturbs cosmic rays to reach the ground). Atmospheric pressure at the time was later checked with the data from automated meteorological data acquisition system (AMeDAS) of Japan Meteorological Agency.

3. *The team of radiation dosimetry under the direct control of Dean, School of Medicine*

The results of surveillance by the team of radiation dosimetry under the direct control of Dean, School of Medicine, are informed to FMU staff twice a month nowadays. Indoor levels are now no problem, while outdoor values are a little high, especially on some 'hotspots.' However, times for club activities outside are not so long, and they can be careful not to stay too long near those hotspots.

**DISCUSSION**

From the early stage of the increase of gamma-ray background caused by the crisis in Fukushima Daiichi nuclear power plant, the author distributed the data as Figure 1 to the relevant people within the campus almost every day, to confirm that there was no further accident in the nuclear power plant. Frequency of the data distribution became roughly once a month nowadays. The audience of this information asked the author several questions.

"How could we translate the unit cps in Figure 1 to more familiar unit of μSv/h?" This was a serious but quite difficult problem. The author had no high precision dosimeters for environmental level. Only a pocket dosimeter (Panasonic, ZP-145) was placed in the author's office from the 18th of March, and comparing cps of NaI with this tiny dosimeter, a conversion factor of $9.4 \times 10^{-4}$ (μSv/h)/cps was temporarily obtained. The author intends to get more precise dosimeter.

"The reason of the increase of long half-life is the influence of supplies from nuclear reactor?" Concerning this question, there is an interesting report by Meteorological Research Institute[4]. This report says that the fallout of Cs-137 away from the Chernobyl nuclear disaster faded environmentally with the half-life of 25 days. Thus, the present author guessed that the newly supplied radioisotopes from Fukushima Daiichi Nuclear Power Station might be the origin of the half-life longer than 25 days. Nowadays, however, environmentally accumulated long HL radioisotopes such as Cs-137 might be the main origin of the apparent half-life longer than 100 days.

"Is the expression with half-life really appropriate for the time change of cps?" The answer is not so clear either. However, radioisotopes from nuclear power plant are divided into two groups, i.e., with short half-life one (I-131) and with long half-life ones (Cs-134, Cs-137 and Sr-90). Therefore, the idea of the sum of two groups of RI with different half-life may be one of the simplest models, at the present time of three months or 100 days after the accident.

## CONCLUSIONS

Three groups of radiation survey through Tohoku Region Pacific Coast Earthquake and the subsequent Fukushima Daiichi nuclear disaster were reported. First, at the head of Associate

Dean, School of Medicine, the group of interested people assembled from the faculty members of "Life Sciences and Social Medicine" and "Human and Natural Sciences" began radiation surveillance immediately after the earthquake, and gave precious information and confirmation of a sense of security for the staff of Fukushima Medical University. Second, serial measurements of natural radiation revealed various responses from the nuclear power plant accidents. For the gamma radiation data, non-linear least squares fit indicated short and long half-life decrease of the radiation. Shorter half-life is clearly recognized as the contribution of iodine 131, while longer half-life is attributed to the radiation from cesium-134, cesium 137 and strontium 90. Third, the team of radiation dosimetry under the direct control of Dean, School of Medicine, started late April and continues the surveillance and will continue for the all the people in Fukushima Medical University.

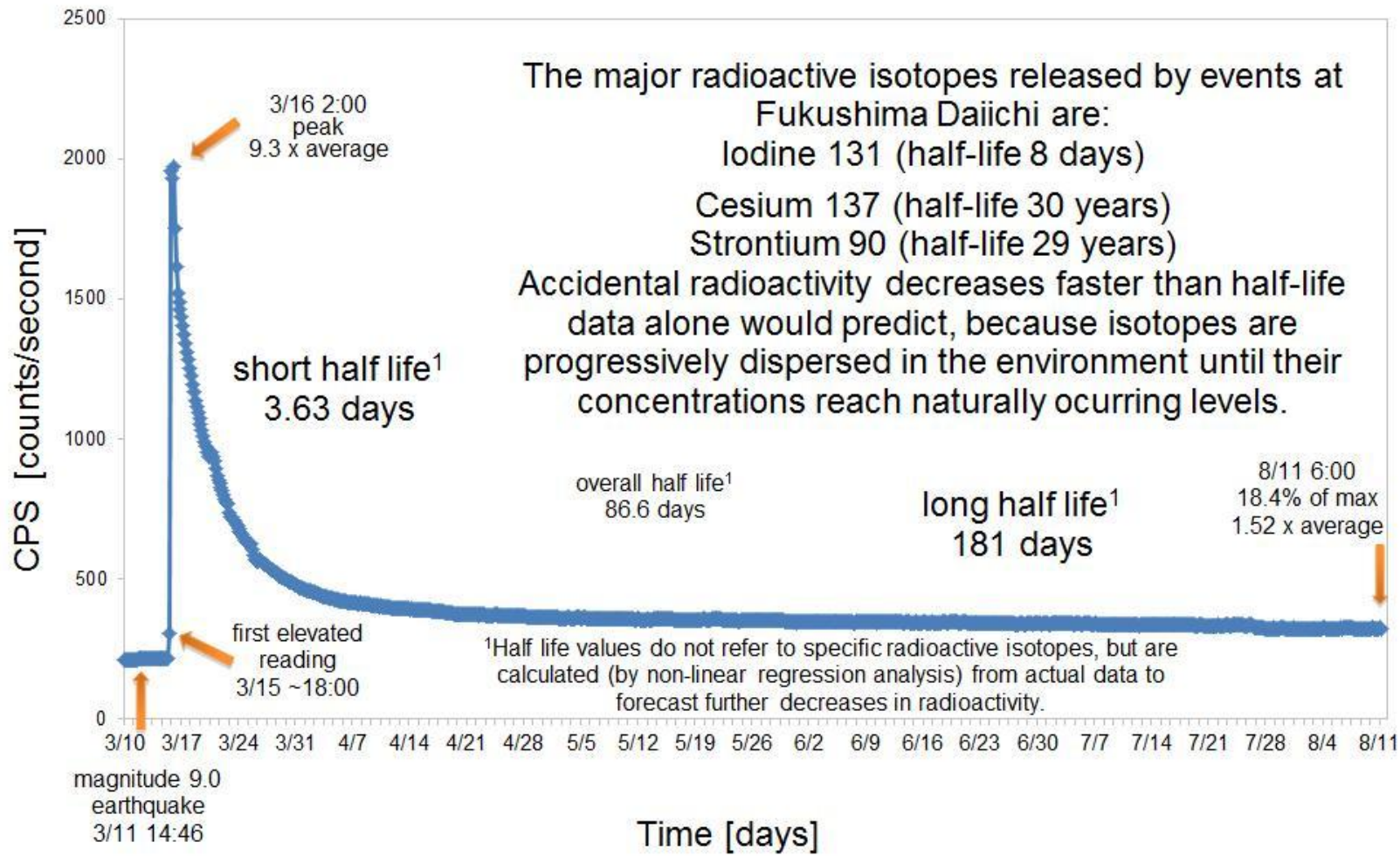

Fig. 1. Serial gamma radiation measurements in air with sodium iodide (NaI) scintillation counter before and after the magnitude 9.0 earthquake, tsunami, and subsequent nuclear power plant crisis. Results were stored every 4 hours as the average cps from the accumulated counts. Measured place was Fukushima Medical University Department of Natural Sciences (Physics) professor's office. Radiation surveillance is a routine activity of this department.

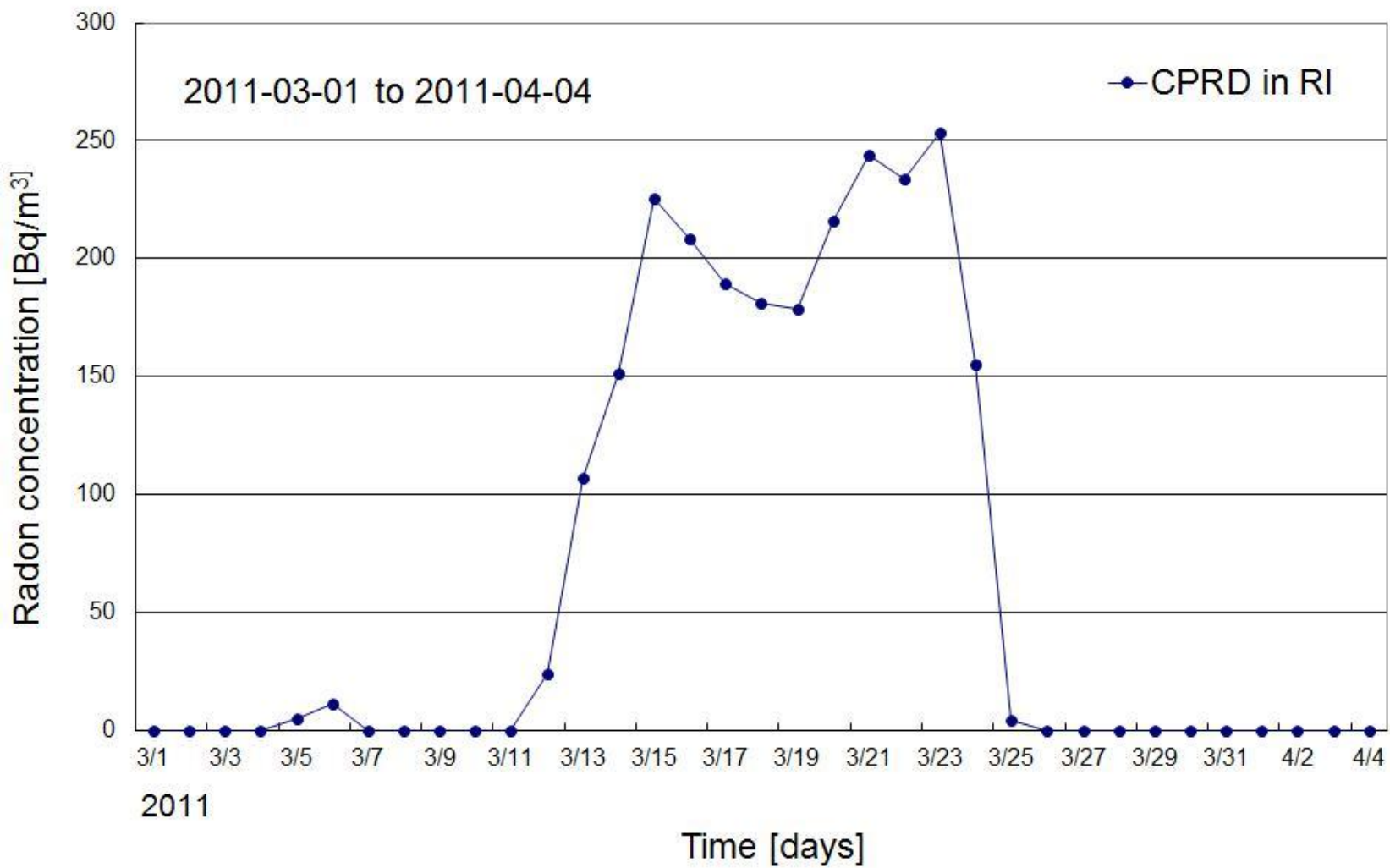

Fig. 2. Change of atmospheric radon concentration after the magnitude 9.0 earthquake in a room of the Radioisotope Center. CPRD means continuous passive radon detector (Pylon, AB-5).